\documentclass[10pt,twoside,twocolumn]{IEEEtran}%
\usepackage{latexsym}
\usepackage{amsmath}
\usepackage{graphicx}
\usepackage{epsfig}
\usepackage{amsfonts}
\usepackage{amssymb}%
\newcommand{\nc}{\newcommand}
\nc{\nn}{\nonumber} \nc{\ep}{\varepsilon} \nc{\la}{\lambda}
\nc{\wht}{\widehat} \nc{\ov}{\overline} \nc{\ds}{\displaystyle}
\nc{\kro}{\left(}\nc{\kvo}{\left[}\nc{\fio}{\left\{}
\nc{\krz}{\right)}\nc{\kvz}{\right]}\nc{\fiz}{\right\}}

\begin{document}

\title{Soft decision decoding of Reed-Muller codes: recursive lists}
\date{}
\author{Ilya Dumer and Kirill Shabunov \thanks{Ilya Dumer is with the College of
Engineering, University of California, Riverside, CA 92521; e-mail:
dumer@ee.ucr.edu. Kirill Shabunov is with the XVD Corporation, San Jose, CA,
95134; e-mail: kshabunov@xvdcorp.com. This research was supported by NSF grant
CCR-0097125.}}
\maketitle

\begin{abstract}
Recursive list decoding is considered for Reed-Muller (RM) codes. The
algorithm repeatedly relegates itself to the shorter RM codes by recalculating
the posterior probabilities of their symbols. Intermediate decodings are only
performed when these recalculations reach the trivial RM codes. In turn, the
updated lists of most plausible codewords are used in subsequent decodings.
The algorithm is further improved by using permutation techniques on code
positions and by eliminating the most error-prone information bits. Simulation
results show that for all RM codes of length 256 and many subcodes of length
512, these algorithms approach maximum-likelihood (ML) performance within a
margin of 0.1 dB. As a result, we present tight experimental bounds on ML
performance for these codes.


\end{abstract}

\section{Introduction}

The main goal of this paper is to design feasible error-correcting algorithms
that approach ML decoding on the moderate lengths ranging from 100 to 1000
bits. The problem is practically important due to the void left on these
lengths by the best algorithms known to date. In particular, exact ML decoding
has huge decoding complexity even on the blocks of 100 bits. On the other
hand, currently known \ iterative (message-passing) algorithms have been
efficient only on the blocks of thousands of bits.

To achieve near-ML performance with moderate complexity, \ we wish to use
\textit{recursive} techniques that repeatedly split an original \ code into
the shorter ones. For this reason, we consider Reed-Muller (RM) codes, which
represent the most notable example of recursive constructions known to date.
These codes - denoted below $\left\{
%TCIMACRO{\QATOP{m}{r}}%
%BeginExpansion
\genfrac{}{}{0pt}{}{m}{r}%
%EndExpansion
\right\}  $ - have length $n=2^{m}$ and Hamming distance $d=2^{m-r}.$ They
also admit a simple recursive structure based on the \textit{Plotkin
construction }$(\mathbf{u,u+v}),$ which splits the original RM code into the
two shorter codes of length $2^{m-1}$. This structure was efficiently used in
recursive decoding algorithms of \cite{lit}-\cite{kab}, which derive the
corrupted symbols of the shorter codes $\mathbf{u}$ and $\mathbf{v}$ from the
received symbols. These recalculations are then repeated until the process
reaches the repetition codes or full spaces, whereupon new information symbols
can be retrieved by any powerful algorithm - say, ML decoding. As a result,
recursive algorithms achieve bounded distance decoding \ with a low complexity
order of $n\min\{r,m-r\}$, which improves upon the complexity of majority
decoding \cite{ree}.

We also mention two list decoding algorithms of \cite{bos1} and \cite{sor1},
which substantially reduce the error rates at the expense of a higher
complexity. In both algorithms, RM codes are represented as the generalized
concatenated codes, which are repeatedly decomposed into the shorter blocks
similarly to the Plotkin construction. In all intermediate steps, the
algorithm of \cite{bos1} tries to estimate the Euclidean distance to the
received vector and then retrieves the codewords with the smallest estimates.
To do so, the algorithm chooses some number $L$ of codewords from both
constituent codes $\mathbf{u}$ and $\mathbf{v.}$ Then the product list is
constructed for the original code. These lists are recursively re-evaluated
and updated in \textit{multiple runs}. The second technique \cite{sor1} is
based on a novel sequential scheme that uses both the main stack and the
complementary one. The idea here is to lower-bound the minimum distance
between the received vector and the best code candidates that will be obtained
in the \textit{future steps}. This ``look-ahead'' approach gives low error
rates and reduces the decoding complexity of \cite{bos1}.

Recently, new recursive algorithms were considered in \cite{dum5} and
\cite{sha}. In particular, for long RM codes of fixed code rate $R,$ recursive
decoding of \cite{dum5} corrects most error patterns of weight ($d\ln d)/2$
instead of the former threshold of $d/2.$ This is done without any increase in
decoding complexity. However, the new decoding threshold is still inferior to
that of a much more powerful ML decoding.

In the sequel, we advance the algorithm of \cite{dum5}, also applying list
decoding techniques. This approach mostly follows \cite{sha} and differs from
the prior results in a few important aspects. First, we use exact posterior
probabilities in our recursive recalculations instead of the distance
approximations employed before. This allows us to design a tree-like recursive
algorithm that can better sort out all plausible candidates in intermediate
steps and avoid multiple decoding runs. Second, \ we\ shall see that the
output error rate significantly varies for the different information symbols
derived in the recursive process. Therefore, we also consider subcodes of RM
codes obtained by removing the least protected information bits. Finally,
decoding will be improved by applying a few permutations on code positions. As
a result, we closely approach the performance of ML decoding on the lengths
256 and 512, which was beyond the reach of the former techniques.

The material is organized as follows. \ In Section 2, we briefly summarize
some recursive properties of RM codes and their decoding procedures. In
Section 3, we describe our list decoding algorithm $\Psi_{\,r}^{m}(L)$.
Finally, in Section 4 we discuss the improvements obtained by eliminating the
least protected information bits and using permutation techniques.

\section{Recursive encoding and decoding for RM codes}

\subsection{Encoding}

The following description is detailed in \cite{dum6}. Let any codeword
$\mathbf{c}$ of RM code $\left\{
%TCIMACRO{\QATOP{m}{r}}%
%BeginExpansion
\genfrac{}{}{0pt}{}{m}{r}%
%EndExpansion
\right\}  $ be represented in the form $\mathbf{u,u+v}$ where $\mathbf{u}%
\in\left\{
%TCIMACRO{\QATOP{m-1}{r}}%
%BeginExpansion
\genfrac{}{}{0pt}{}{m-1}{r}%
%EndExpansion
\right\}  $ and $\mathbf{v\in}\left\{
%TCIMACRO{\QATOP{m-1}{r-1}}%
%BeginExpansion
\genfrac{}{}{0pt}{}{m-1}{r-1}%
%EndExpansion
\right\}  $. We say that $\mathbf{c}$ is split onto two ``paths'' $\mathbf{u}$
and $\mathbf{v}$. By splitting both paths, \ we obtain four paths that lead to
RM codes of length $2^{m-2},$ and so on. In each step $i$ of our splitting, we
assign the path value $\xi_{i}=0$ to a new $\mathbf{v}$-component and $\xi
_{i}=1$ to a new $\mathbf{u}$-component. \ All paths end at the repetition
codes $\left\{
%TCIMACRO{\QATOP{g}{0}}%
%BeginExpansion
\genfrac{}{}{0pt}{}{g}{0}%
%EndExpansion
\right\}  $ or full spaces $\left\{
%TCIMACRO{\QATOP{h}{h}}%
%BeginExpansion
\genfrac{}{}{0pt}{}{h}{h}%
%EndExpansion
\right\}  ,$ where
\[
g=1,...,m-r,\ \quad h=1,...,r.
\]
Thus, we can consider a specific binary path%
\[
\xi\overset{\text{def}}{=}(\xi_{1},...,\xi_{m-g})
\]
that leads from the origin $\left\{
%TCIMACRO{\QATOP{m}{r}}%
%BeginExpansion
\genfrac{}{}{0pt}{}{m}{r}%
%EndExpansion
\right\}  $ to some left-end code $\left\{
%TCIMACRO{\QATOP{g}{0}}%
%BeginExpansion
\genfrac{}{}{0pt}{}{g}{0}%
%EndExpansion
\right\}  $. For any right-end node $\left\{
%TCIMACRO{\QATOP{h}{h}}%
%BeginExpansion
\genfrac{}{}{0pt}{}{h}{h}%
%EndExpansion
\right\}  ,$ the same process gives a subpath $\xi$ of length $m-h:$
\[
\xi\overset{\text{def}}{=}(\xi_{1},...,\xi_{m-h}).
\]

A similar decomposition can be performed on the block $\mathbf{a}_{\,r}^{m}$
of \ $k\ $information bits \ that encode the original vector $\mathbf{c}.$ In
this way, any left-end path $\xi$ gives only one information bit associated
with its end node $\left\{
%TCIMACRO{\QATOP{g}{0}}%
%BeginExpansion
\genfrac{}{}{0pt}{}{g}{0}%
%EndExpansion
\right\}  .$ Any right-end path gives $2^{h}$ information bits associated with
the end code $\left\{
%TCIMACRO{\QATOP{h}{h}}%
%BeginExpansion
\genfrac{}{}{0pt}{}{h}{h}%
%EndExpansion
\right\}  .$ We can also add an arbitrary binary suffix of length $h$ to the
right-end paths, and obtain a one-to-one mapping between the extended paths
$\xi$ and $k$ information bits $a(\xi).$

\subsection{Basic decoding with posterior probabilities}

Let any binary symbol $a$ be mapped onto $(-1)^{a}$. Then any codeword of RM
code belongs to $\{1,-1\}^{n}$ and has the form $\mathbf{c=}(\mathbf{u,uv}).$
This codeword is transmitted over a memoryless channel $\mathcal{Z}_{g}.$ The
received block $\mathbf{x}$ consists of the two halves $\mathbf{x}^{\prime}$
and $\mathbf{x}^{\prime\prime}$, which are the corrupted images of vectors
$\mathbf{u}$ and $\mathbf{uv}$. \ The decoder first takes the symbols
$x_{i}^{\prime}$ and $x_{i}^{\prime\prime}$ for any position $i=1,...,n/2,$%
\ and finds the posterior probabilities of transmitted symbols $u_{i}$ and
$u_{i}v_{i}:$
\[
q_{i}^{\prime}\overset{\text{def}}{=}\Pr\{u_{i}=1\,\,|\,\,x_{i}^{\prime
}\},\quad q_{i}^{\prime\prime}\overset{\text{def}}{=}\Pr\{u_{i}v_{i}%
=1\,\,|\,\,x_{i}^{\prime\prime}\}.\smallskip
\]
To simplify our notation, below we use the associated quantities
\begin{equation}
y_{i}^{\prime}\overset{\text{def}}{=}2q_{i}^{\prime}-1,\qquad y_{i}%
^{\prime\prime}=2q_{i}^{\prime\prime}-1. \label{pe}%
\end{equation}
Note that $y_{i}^{\prime}$ is the \textit{difference }between the two
posterior probabilities $q_{i}^{\prime}$ and $1-q_{i}^{\prime}$ of $1$ and
$-1$ in position $i$ of the left half. Similarly, $y_{i}^{\prime\prime}$ is
obtained on the right half. The following basic recursive algorithm is
described in \cite{dum5} and Section IV of \cite{dum6} in more detail.

\textbf{Step 1. }Let $q_{i}^{v}=\Pr\{v_{i}=1\,\,|\,\,x_{i}^{\prime}%
\,,x_{i}^{\prime\prime}\}$ be the posterior probability of any symbol
$v_{i}\mathbf{\ }$of \ the codeword $\mathbf{v.}$ We find the corresponding
quantity $y_{i}^{v}=2q_{i}^{v}-1,$ which is (see formula (18) in \cite{dum6})
\begin{equation}
y_{i}^{v}=y_{i}^{\prime}y_{i}^{\prime\prime}. \label{1}%
\end{equation}
Symbols $y_{i}^{v}$ form the vector $\mathbf{y}^{v}$ of length $n/2.$ Then we
use some soft-decision decoder $\Psi_{v}(\mathbf{y}^{v})$ that gives a vector
$\hat{\mathbf{v}}\in\left\{
%TCIMACRO{\QATOP{m-1}{r-1}}%
%BeginExpansion
\genfrac{}{}{0pt}{}{m-1}{r-1}%
%EndExpansion
\right\}  $ and its information block $\hat{\mathbf{a}}^{v}.$

\textbf{Step 2. }Now we assume that $\ $Step 1 gives\textit{ correct}
\textit{vector} $\hat{\mathbf{v}}=\mathbf{v.}$ Let $q_{i}^{u}=\Pr
\{u_{i}=1\,\,|\,\,x_{i}^{\prime}\,,x_{i}^{\prime\prime}\}$ be the posterior
probability of a symbol $u_{i}.$ Then the corresponding quantity $y_{i}%
^{u}=2q_{i}^{u}-1$ is (see formula (19) in \cite{dum6})
\begin{equation}
y_{i}^{u}=(y_{i}^{\prime}+\hat{y}_{i})/(1+y_{i}^{\prime}\hat{y}_{i}),
\label{2}%
\end{equation}
where $\hat{y}_{i}=y_{i}^{\prime\prime}\hat{v}_{i}.$ The symbols $y_{i}^{u}$
form the vector $\mathbf{y}^{u}$ of length $n/2.$ We use some (soft decision)
decoding algorithm $\Psi_{u}(\mathbf{y}^{u})$ to obtain a vector
$\hat{\mathbf{u}}\mathbf{\in}\left\{
%TCIMACRO{\QATOP{m-1}{r}}%
%BeginExpansion
\genfrac{}{}{0pt}{}{m-1}{r}%
%EndExpansion
\right\}  $ and its information block $\hat{\mathbf{a}}^{u}.$ \hfill$\square
$\smallskip

In a more general scheme $\Psi_{r}^{m}$, vectors $\mathbf{y}^{v}$ and
$\mathbf{y}^{u}$ are not decoded but used as our new inputs $\mathbf{y.}$
These inputs are recalculated multiple times according to (\ref{1}) and
(\ref{2}). Finally, we reach the end nodes $\left\{
%TCIMACRO{\QATOP{g}{0}}%
%BeginExpansion
\genfrac{}{}{0pt}{}{g}{0}%
%EndExpansion
\right\}  $ and $\left\{
%TCIMACRO{\QATOP{h}{h}}%
%BeginExpansion
\genfrac{}{}{0pt}{}{h}{h}%
%EndExpansion
\right\}  $. Here we perform maximum-likelihood (ML) decoding as follows.

At any node $\left\{  _{h}^{g}\right\}  $, our input is a newly recalculated
vector $\mathbf{y}$ of length $2^{g}$ with the given differences $y_{i}$
between posterior probabilities of two symbols $c_{i}=\pm1$. Rewriting
definition (\ref{pe}), we assign the posterior probability
\[
\Pr(c_{i}|y_{i}\mathbf{)=}(1+c_{i}y_{i})/2
\]
to a symbol $c_{i}=\pm1$. In this way, we can find the posterior probability
\begin{equation}
P(\mathbf{c}\mid\mathbf{y)=}\prod_{i=1}^{2^{g}}(1+c_{i}y_{i})/2 \label{ml1}%
\end{equation}
of any codeword $\mathbf{c\in}\left\{  _{h}^{g}\right\}  $, and choose the
most probable codeword $\mathbf{\hat{c}}$, where
\begin{equation}
\forall\mathbf{c}\in\left\{  _{h}^{g}\right\}  :P(\mathbf{\hat{c}%
}|\mathbf{y)\geq}P\mathbf{(\mathbf{c}|y\mathbf{)}.} \label{ml2}%
\end{equation}
The decoded codeword~$\mathbf{\hat{c}\in}\left\{  _{r}^{m}\right\}  $ and the
corresponding information block~$\hat{\mathbf{a}}$ are now obtained as follows
(here operations (\ref{1}) and (\ref{2}) are performed on vectors componentwise).%

\[
\frame{$%
\begin{array}
[c]{l}%
\text{Algorithm }\Psi_{r}^{m}\text{ for an input vector }\mathbf{y.}\medskip\\
\text{1. If }0<r<m\text{, execute the following.}\medskip\\%
\begin{array}
[c]{l}%
\quad\text{1.1.\ Calculate vector }\mathbf{y}^{v}=\mathbf{y}^{\prime
}\mathbf{y}^{\prime\prime}\text{.}\smallskip\text{\smallskip}\\
\quad\text{Decode }\mathbf{y}^{v}\text{ into vector }\mathbf{\hat{v}=}%
\Psi_{\text{ }r-1}^{m-1}(\mathbf{y}^{v}).\smallskip\text{ }\\
\quad\text{Pass }\mathbf{\hat{v}}\text{ and }\hat{\mathbf{a}}^{v}\text{ to
Step 1.2}\medskip\\
\quad\text{1.2.\ Calculate vector }\mathbf{y}^{u}=(\mathbf{y}^{\prime
}+\mathbf{\hat{y}})/(1+\mathbf{y}^{\prime}\mathbf{\hat{y}})\text{.\smallskip
}\smallskip\\
\quad\text{Decode }\mathbf{y}^{u}\text{ into vector }\mathbf{\hat{u}=}\Psi
_{r}^{m-1}(\mathbf{y}^{u}).\smallskip\ \\
\quad\text{Output decoded components:}\\
\quad\hat{\mathbf{a}}:=(\hat{\mathbf{a}}^{v}\mid\hat{\mathbf{a}}^{u}%
);\quad\mathbf{\hat{c}}:=(\mathbf{\hat{u}}\mid\mathbf{\hat{u}\hat{v}%
}).\medskip
\end{array}
\\
\text{2. If }r=0,\text{ use ML-decoding (\ref{ml2})\ for \ }\left\{
%TCIMACRO{\QATOP{r}{0}}%
%BeginExpansion
\genfrac{}{}{0pt}{}{r}{0}%
%EndExpansion
\right\}  .\medskip\\
\text{3. If }r=m,\text{ use ML-decoding (\ref{ml2}) for }\left\{
%TCIMACRO{\QATOP{r}{r}}%
%BeginExpansion
\genfrac{}{}{0pt}{}{r}{r}%
%EndExpansion
\right\}  .\smallskip
\end{array}
$}%
\]
Note that this algorithm $\Psi_{\,r}^{m}$ differs from the simplified
algorithm $\Phi_{\,r}^{m}$ of \cite{dum6} in three aspects. Firstly, we use
exact recalculations (\ref{2}) instead of the former simplification%
\begin{equation}
\mathbf{y}^{u}=(\mathbf{y}^{\prime}+\mathbf{\hat{y}})/2. \label{2-1}%
\end{equation}
Secondly, we use ML decoding instead of the minimum distance decoding that
chooses $\mathbf{\hat{c}}$ with the maximum inner product:
\[
\forall\mathbf{c}:(\mathbf{\hat{c}},\mathbf{y)\geq(\mathbf{c,}y\mathbf{)}.}%
\]
Thirdly, we employ a different rule and stop at the repetition codes $\left\{
%
%TCIMACRO{\QATOP{r}{0}}%
%BeginExpansion
\genfrac{}{}{0pt}{}{r}{0}%
%EndExpansion
\right\}  $ instead of the biorthogonal codes used in \cite{dum6}. This last
change will make it easier to use the list decoding described in the following section.

Finally, note that recalculations (\ref{1}) require $n/2$ operations, while
recalculations (\ref{2}) can be done \ in $5n/2$ operations. Therefore our
decoding complexity satisfies recursion%
\[
|\Psi_{\,r}^{m}|\leq\left|  \Psi_{\text{ }r-1}^{m-1}\right|  +\left|
\Psi_{\,\,\,\,r}^{m-1}\right|  +3n.
\]
Similarly to \cite{dum6}, this recursion gives decoding complexity
\[
|\Psi_{\,r}^{m}|\leq6n\min(r,m-r)+n.
\]
Thus, complexity $|\Psi_{\,r}^{m}|$ has maximum order of $3n\log n,$ which is
twice the complexity $\left|  \Phi_{\,r}^{m}\right|  $ of the algorithm
$\Phi_{\,r}^{m}$ of \cite{dum6}.

\section{List decoding}

To enhance algorithm $\Psi_{r}^{m}$, we shall use some lists of $L=2^{p}$ or
fewer codewords obtained on any path $\xi$. This algorithm -- called $\Psi
_{r}^{m}(L)$ -- increases the number of operations at most $L$ times and has
the overall complexity order of $Ln\log n.$ Given any integer parameter $A$,
\ we say that the list have size $A^{\ast},$ if decoding outputs either all
available records or $A$ records, whichever is less$.$ This algorithm performs
as follows.

At any step $s=1,...,k$ of the algorithm $\Psi_{r}^{m}(L)$, our input consists
of $L^{\ast}$ records
\[
A=(\mathbf{\bar{a}},\rho(\mathbf{\bar{a}}),\mathbf{y(\bar{a}})).
\]
Each record is formed by some information block $\mathbf{\bar{a}},$ its cost
function $\rho(\mathbf{\bar{a}}),$ and the corresponding input $\mathbf{y(\bar
{a}}),$ which is updated in the decoding process These three entries are
defined below.

Decoding\textit{ }starts at the root node $\left\{
%TCIMACRO{\QATOP{m}{r}}%
%BeginExpansion
\genfrac{}{}{0pt}{}{m}{r}%
%EndExpansion
\right\}  .$ Here we set $s=0$ and take one record
\begin{equation}
\mathbf{\bar{a}}=\emptyset,\quad\rho(\mathbf{\bar{a}})=1,\quad\mathbf{y(\bar
{a}})=\mathbf{y,} \label{ini1}%
\end{equation}
where $\mathbf{y}$ is the input vector. Decoding takes the first path (denoted
$\xi=1)$ to the leftmost code $\left\{
%TCIMACRO{\QATOP{m-r}{0}}%
%BeginExpansion
\genfrac{}{}{0pt}{}{m-r}{0}%
%EndExpansion
\right\}  $ and recalculates vector $\mathbf{y(\bar{a}})$ similarly to the
algorithm $\Psi_{r}^{m}.$ However, now we take \textit{both} values
$a_{1}=0,1$ of \ the first information symbol and consider both codewords
$1^{d}$ and $-1^{d}$ of length $d=2^{m-r}$ in the repetition code
$\mathbf{c(}a_{1})$. The posterior probabilities (\ref{ml1}) of these two
vectors will also define the cost function of the new information block
$\mathbf{\bar{a}}=a_{1}:$
\[
\rho(\mathbf{\bar{a}})\mathbf{=}\prod_{i=1}^{2^{m-r}}\frac{1+\mathbf{c}%
_{i}(a_{1})y_{i}\mathbf{(\bar{a}})}{2}%
\]
In our list decoding, we represent the two outcomes $\mathbf{\bar{a}}$ as the
initial edges mapped with their cost functions $P(\mathbf{\bar{a}})\mathbf{.}$
Then we proceed to the next code $\left\{
%TCIMACRO{\QATOP{m-r-1}{0}}%
%BeginExpansion
\genfrac{}{}{0pt}{}{m-r-1}{0}%
%EndExpansion
\right\}  ,$ which corresponds to the subsequent path denoted $\xi=2.$ Given
two different decoding results $\mathbf{v}=\mathbf{c(}a_{1}),\mathbf{\ }$our
recursion (\ref{1}), (\ref{2}) gives two different vectors $\mathbf{y(\bar{a}%
})$ arriving at this node. \ Therefore, decoding is performed two times and
gives the full tree of depth 2. More generally, at any step $s$, decoding is
executed as follows.

Suppose that the first $s-1$ paths are already processed. This gives $L^{\ast
}$ information blocks
\[
\mathbf{\bar{a}}=(a_{1},...,a_{s-1})
\]
of length $s-1$ and the corresponding records $A.$ Each vector $\mathbf{y(\bar
{a}})$ is then recalculated on the new path $\xi=s$ using formulas (\ref{1})
and (\ref{2}), in the same way it was done in $\Psi_{r}^{m}.$ Let this path
end on some left-end code $\left\{
%TCIMACRO{\QATOP{g}{0}}%
%BeginExpansion
\genfrac{}{}{0pt}{}{g}{0}%
%EndExpansion
\right\}  $. Decoding of the new information symbol $a_{s}=0,1$ yields
$2L^{\ast}$ extended blocks
\[
\mathbf{\bar{a}}:=\mathbf{\bar{a}},a_{s}%
\]
of depth $s,$ marked by their cost functions%
\begin{equation}
\rho(\mathbf{\bar{a}}):=\rho(\mathbf{\bar{a}})\cdot\prod_{i=1}^{2^{g}}%
\frac{1+c_{i}(a_{s})y_{i}(\mathbf{\bar{a}})}{2}. \label{cost}%
\end{equation}
Step $s$ is completed by choosing $L^{\ast}$ blocks with the highest cost
functions $\rho(\mathbf{\bar{a}}).$

The decoding on the right-end nodes is similar. The only difference is that
the full spaces $\left\{  _{h}^{h}\right\}  $ include $2^{h}$ codewords
defined by information blocks $a_{s}$ of length $|a_{s}|=h.$ In this case, we
can choose the two most probable vectors $\mathbf{c}(a_{s})$ (in essence,
making bit-by-bit decisions) and set $g=h$ in our cost calculations
(\ref{cost}). Another - more refined version of the algorithm - chooses four
different vectors of the code $\left\{  _{h}^{h}\right\}  $ whenever $h\geq2.$
The best record is chosen at the last node $\left\{
%TCIMACRO{\QATOP{r}{r}}%
%BeginExpansion
\genfrac{}{}{0pt}{}{r}{r}%
%EndExpansion
\right\}  .$ More generally, the algorithm is executed as follows.%

\[
\frame{$%
\begin{array}
[c]{l}%
\text{Algorithm }\Psi_{r}^{m}(L).\text{ Input: }L^{\ast}\text{ records}\\
\text{ }A=(\mathbf{\bar{a}},\rho(\mathbf{\bar{a}}),\mathbf{y(\bar{a}})),\text{
counter }s=0.\medskip\\
\text{1. If }0<r<m\text{, for all vectors }\mathbf{y(\bar{a}}):\medskip\\%
\begin{array}
[c]{l}%
\quad\text{1.1. Set }\mathbf{y(\bar{a}}):=\mathbf{y}^{\prime}\mathbf{(\bar{a}%
})\mathbf{y}^{\prime\prime}\mathbf{(\bar{a}})\text{.}\medskip\\
\quad\text{Perform decoding }\Psi_{\text{ }r-1}^{m-1}\left(  \mathbf{y(\bar
{a}})\right)  .\medskip\\
\quad\text{Pass }L^{\ast}\text{ new records }A\text{ to Step 1.2}\medskip\\
\quad\text{1.2.\ Set }\mathbf{y(\bar{a}}):=\ds\frac{\mathbf{y}^{\prime
}\mathbf{(\bar{a}})+\mathbf{\hat{y}(\bar{a}})}{1+\mathbf{y}^{\prime
}\mathbf{(\bar{a}})\mathbf{\hat{y}(\bar{a}})}\text{.}\medskip\smallskip\\
\quad\text{Perform decoding }\Psi_{r}^{m-1}\left(  \mathbf{y(\bar{a}})\right)
.\smallskip\ \\
\quad\text{Output }L^{\ast}\text{ new records }A\text{.\smallskip}\smallskip
\end{array}
\\
\text{2. If }r=0,\text{ take both values }a_{s}=0,1.\\
\text{Calculate costs (\ref{cost}) for each }\left(  \mathbf{\bar{a},}%
a_{s}\right)  .\smallskip\\
\text{Choose }L^{\ast}\text{ best blocks }\mathbf{\bar{a}:=(\bar{a},}a_{s}).\\
\text{Set }s:=s+1\text{ and return }L^{\ast}\text{ records }A.\medskip\\
\text{3. If }r=m,\text{ choose 4}^{\ast}\text{ best blocks }a_{s}.\\
\text{Calculate costs (\ref{cost}) for each }\left(  \mathbf{\bar{a},}%
a_{s}\right)  \mathbf{.}\smallskip\\
\text{Choose }L^{\ast}\text{ best blocks }\mathbf{\bar{a}:=(\bar{a},}a_{s}).\\
\text{Set }s:=s+|a_{s}|\text{ and return }L^{\ast}\text{ records }A.\medskip
\end{array}
$}%
\]

\textit{Discussion.} \ Consider the above algorithm on the AWGN channel
$\mathcal{N}(0,\sigma^{2}).$ Using the results of \cite{dum6}, it can be
proven that on this channel, the $\mathbf{v}$-component is decoded on the
channel with the new noise power%
\[
\sigma_{v}^{2}\geq\max\{2\sigma^{2},\sigma^{4}\}.
\]
The first approximation is tight for very small $\sigma^{2}$ (though the
channel is no longer Gaussian), while the second one performs well on the
``bad'' channels with $\sigma^{2}\gg1.$ Thus, the noise power always increases
in the $\mathbf{v}$-direction; the more so the worse the original channel is.
By contrast, the $\mathbf{u}$-channel can be approximated by the smaller power
$\sigma^{2}/2$. These observations also show that the first information symbol
- which is obtained on the binary path $0^{r}$ - is protected the least, and
then the decoding gradually improves on the subsequent paths.

Now we see that the algorithm $\Psi_{r}^{m}(L)$ with the list of size
$L=2^{p}$ delays our decision on any information symbol by $p$ steps, making
this decision better protected. In the particular case of a bad channel, it
can be verified that the first symbol $a_{1}$ is now decoded when the noise
power is reduced $2^{p}$ times. For this reason, this list decoding
substantially reduces the output word error rates (WER) even for small size
$L.$

For $L=2^{m-r+1},$ the algorithm $\Psi_{r}^{m}(L)$ processes all the codewords
of the first biorthogonal code $\left\{
%TCIMACRO{\QATOP{m-r+1}{1}}%
%BeginExpansion
\genfrac{}{}{0pt}{}{m-r+1}{1}%
%EndExpansion
\right\}  $ and is similar to the algorithm $\Phi_{\,r}^{m}$ of \cite{dum6}.
\ On the other \ hand, algorithm $\Psi_{r}^{m}(L)$ updates all $L$ cost
functions, while $\Phi_{\,r}^{m}$ chooses one codeword on each end node.
Therefore $\Psi_{r}^{m}(L)$ can be considered as a generalization of
$\Phi_{\,r}^{m}$ that continuously updates decoding lists in all intermediate
steps. The result is a more powerful decoding that comes along with a higher complexity.

\textit{Simulation results. }Below we present our simulation results for the
AWGN channels. Here we also counted all the instances, when for a given output
the decoded vector \ was more probable than the transmitted one. Obviously,
these specific events also represent the errors of ML decoding. Thus, the
fraction of these events gives a lower bound on the ML decoding error
probability. This lower bound is also depicted in the subsequent figures for
all the codes tested.

\begin{figure}[tbh]
\begin{center}
\includegraphics[width=3.5in]{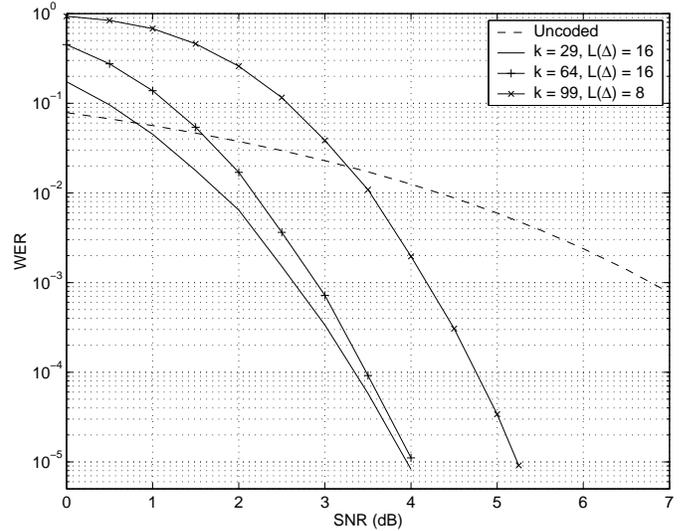}
\end{center}
\caption{Tight lower bounds on WER of ML decoding for three RM codes of length
128. The legend gives the list size $L(\Delta)$ for which the algorithm
$\Psi_{r}^{m}(L)$ performs withing $\Delta=0.25$ dB from these bounds.}%
\label{fig:rm128}%
\end{figure}

Our simulation results show that for all RM codes of lengths 128 and 256, the
algorithm $\Psi_{r}^{m}(L)$ rapidly approaches ML performance as the list size
$L$ grows. \ For RM codes of length 128 and distance $d>4$, we summarize
\ these results in Fig.~\ref{fig:rm128}. For each RM code, we present
\textit{tight} lower bounds for the error probability of ML decoding. To
measure the efficiency of the algorithm $\Psi_{r}^{m}(L),$ we also exhibit the
actual list size $L(\Delta)$ at which $\Psi_{r}^{m}(L)$ approaches the optimal
ML decoding within a small margin of
\[
\Delta=0.25\text{ dB.}%
\]
This performance loss $\Delta$ is measured at the output word error rate (WER)
$P=10^{-4};$ however, we found little to no difference for all other WER
tested in our simulation. In Table 1, we complement these lists sizes
$L(\Delta)$ with the two other relevant parameters:

$-$ the signal-to-noise ratios (SNR per information bit) at which algorithm
$\Psi_{r}^{m}(L)$ gives the WER $P=10^{-4};$

$-$ the complexity estimates $\left|  \Psi_{r}^{m}(L)\right|  $ counted as the
number of floating point operations.

\medskip

\begin{center}%
\begin{tabular}
[c]{|c|c|c|c|}\hline
RM Code & $%
%TCIMACRO{\QATOPD{\{}{\}}{7}{2}}%
%BeginExpansion
\genfrac{\{}{\}}{0pt}{}{7}{2}%
%EndExpansion
$ & $%
%TCIMACRO{\QATOPD{\{}{\}}{7}{3}}%
%BeginExpansion
\genfrac{\{}{\}}{0pt}{}{7}{3}%
%EndExpansion
$ & $\underset{}{\overset{}{%
%TCIMACRO{\QATOPD{\{}{\}}{7}{4}}%
%BeginExpansion
\genfrac{\{}{\}}{0pt}{}{7}{4}%
%EndExpansion
}}$\\\hline
$\overset{}{\underset{}{\text{List size }L(\Delta)}}$ & 16 & 16 & 8\\\hline
$\overset{}{\underset{}{\text{Complexity}}}\left|  \Psi(L)\right|  $ & 21676 &
33618 & 18226\\\hline
$\overset{}{\underset{}{\text{SNR (dB) at }10^{-4}}}$ & 3.47 & 3.71 &
4.85\\\hline
\end{tabular}
\medskip
\end{center}

Table 1. RM codes of length 128: the list size $L(\Delta),$ decoding
complexity, and the corresponding SNR at which algorithm $\Psi_{r}^{m}(L)$
performs within $\Delta=0.25$ dB from ML decoding at WER $10^{-4}.$\medskip

\begin{figure}[tbh]
\begin{center}
\includegraphics[width=3.5in]{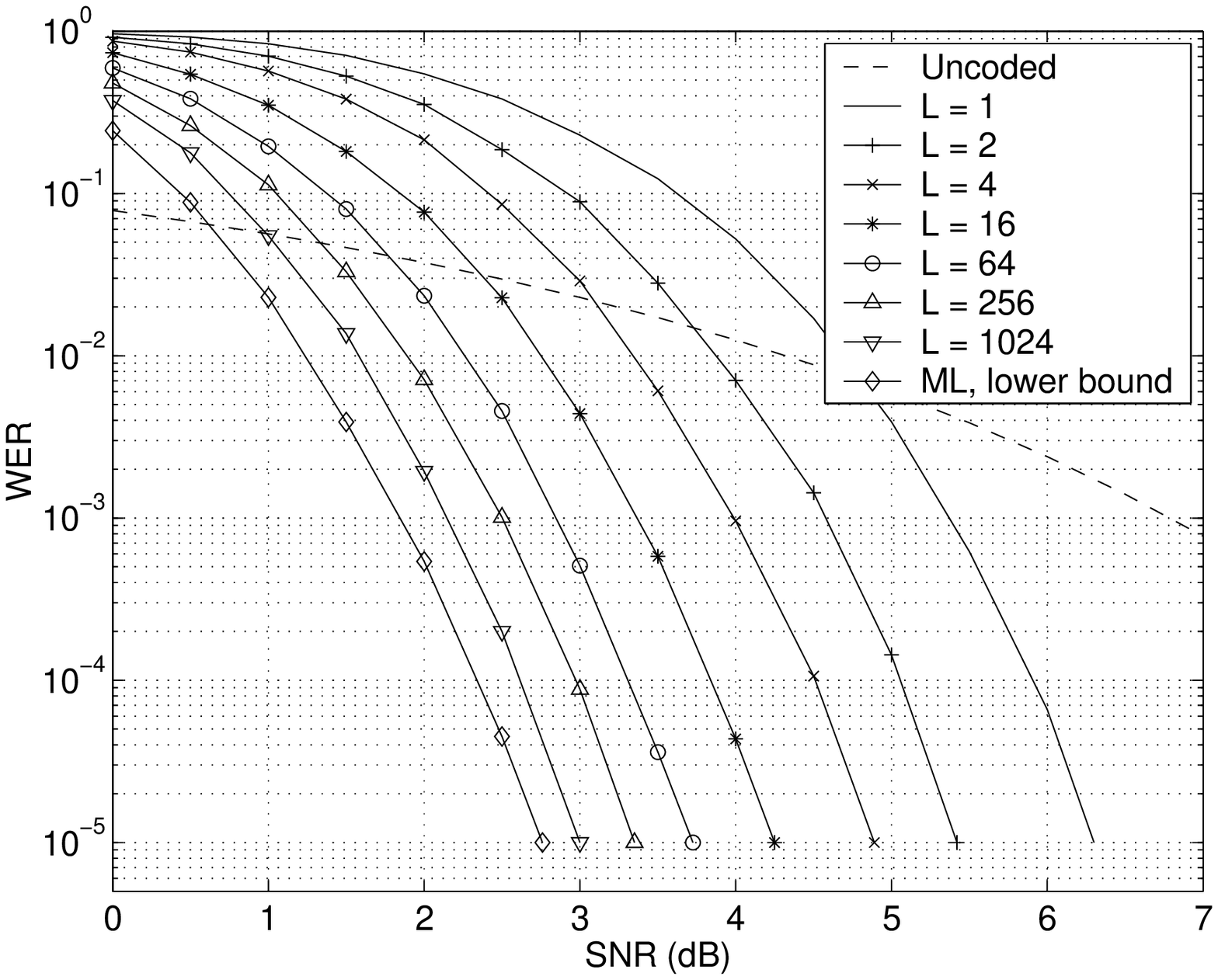}
\end{center}
\caption{ ($256$, $93$) RM code $\genfrac{\{}{\}}{0pt}{}{8}{3}$. WER for the
algorithm $\Psi_{r}^{m}(L)$ with lists of size $L.$}%
\label{fig:rm0308}%
\end{figure}

For RM codes of length 256, we skip most decoding results as these will be
improved in the next section by the permutation techniques. In our single
example in Fig.~\ref{fig:rm0308}, we present the results for the
$(n=256,k=93)$ code $\left\{  _{3}^{8}\right\}  .$ This code gives the
\textit{lowest} rate of convergence to the ML decoding among all RM codes of
length $256.$ In other words, all other codes require the smaller lists to
achieve the same performance loss $\Delta.$ This example and other simulation
results show that the algorithm\textit{ }$\Psi_{r}^{m}(L)$ performs within 0.5
dB from ML decoding on the lengths 128 and 256 using lists of small or
moderate size.

\section{Further improvements}

\subsection{Subcodes of RM codes}

More detailed results also show that many codes of length $n\geq256$ require
lists of large size $L\geq1024$ to approach ML decoding within the small
margin of 0.25 dB. Therefore for $n\geq256$, we also employ a different
approach. Namely, the decoding performance can be improved by eliminating
those paths, where recursive decoding fails more often. Here we use the
results of \cite{dum6}, which show that the leftmost paths are the least protected.

Recall that each left-end path $\xi$ corresponds to one information symbol.
Therefore, decoding on these paths can be eliminated by \textit{setting the
corresponding} \textit{ information bits as zeros}. In this way, we employ the
\textit{subcodes} of the original code $\left\{
%TCIMACRO{\QATOP{m}{r}}%
%BeginExpansion
\genfrac{}{}{0pt}{}{m}{r}%
%EndExpansion
\right\}  $. Note that our decoding algorithm $\Psi_{r}^{m}(L)$ runs virtually
unchanged on subcodes. Indeed, the single difference arises when some
information block $a_{s}$ takes only one value 0 on the corresponding left
node (or less than $2^{h}$ values on the right node). Therefore, on each step
$s,$ we can proceed as before, by taking only the actual blocks $a_{s}$ left
at this node after expurgation.

In the algorithm $\Psi_{r}^{m}(L)$, \ this expurgation starts with the least
protected information path $0^{r}$ that ends at the node $\left\{
%TCIMACRO{\QATOP{m-r}{0}}%
%BeginExpansion
\genfrac{}{}{0pt}{}{m-r}{0}%
%EndExpansion
\right\}  .$ It can be shown that for long RM codes of fixed order $r,$
eliminating even the single weakest path $0^{r}$ \ increases the admissible
noise power $2^{1/2^{r}}$ times$.$ Thus, the lowest orders $r=2,3$ yield the
biggest gain (10log$_{10}2)/2^{r}$ dB, which equals 0.75 dB and 0.375 dB, respectively.

To proceed further, we eliminate the next weakest path $0^{r-1}10.$ However,
the theoretical analysis becomes more complicated on the subsequent bits and
it is unclear which bits should be eliminated first. For this reason, we
optimized this pruning process in our simulation by making a few ad hoc trials
and eliminating subsequent bits in different order.

\begin{figure}[tbh]
\begin{center}
\includegraphics[width=3.5in]{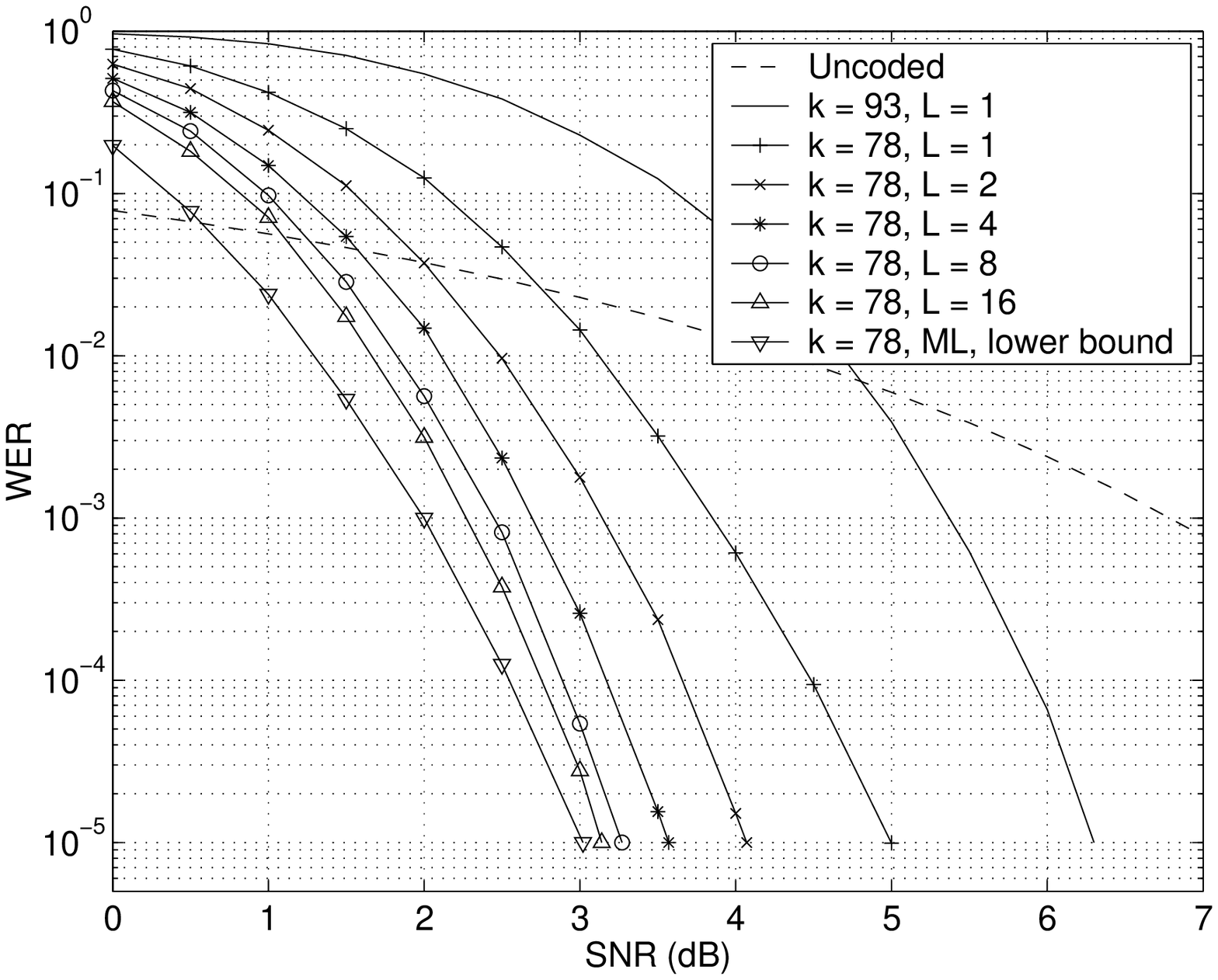}
\end{center}
\caption{($256$, $78$)-subcode of the ($256$, $93$) RM code
$\genfrac{\{}{\}}{0pt}{}{8}{3}.$ WER for the algorithm $\Psi_{r}^{m}(L)$ with
lists of size $L.$}%
\label{fig:srm0308}%
\end{figure}

The corresponding simulation results are presented in Figure~\ref{fig:srm0308}
for the $(256,93)$-code $\left\{  _{3}^{8}\right\}  $ and its $(256,78)$%
-subcode. We see that pruning substantially improves code performance. It is
also interesting to compare Figures~\ref{fig:rm0308} and \ref{fig:srm0308}. We
see that the subcode approaches the optimal ML performance much faster than
the original code does. In particular, the same margin of $\Delta=0.25$ dB can
be reached with only $L=16$ codewords instead of $L=1024$ codewords needed on
the code. In all other examples, the subcodes also demonstrated a much faster
convergence, which leads to a lesser complexity.

\begin{figure}[tbh]
\begin{center}
\includegraphics[width=3.5in]{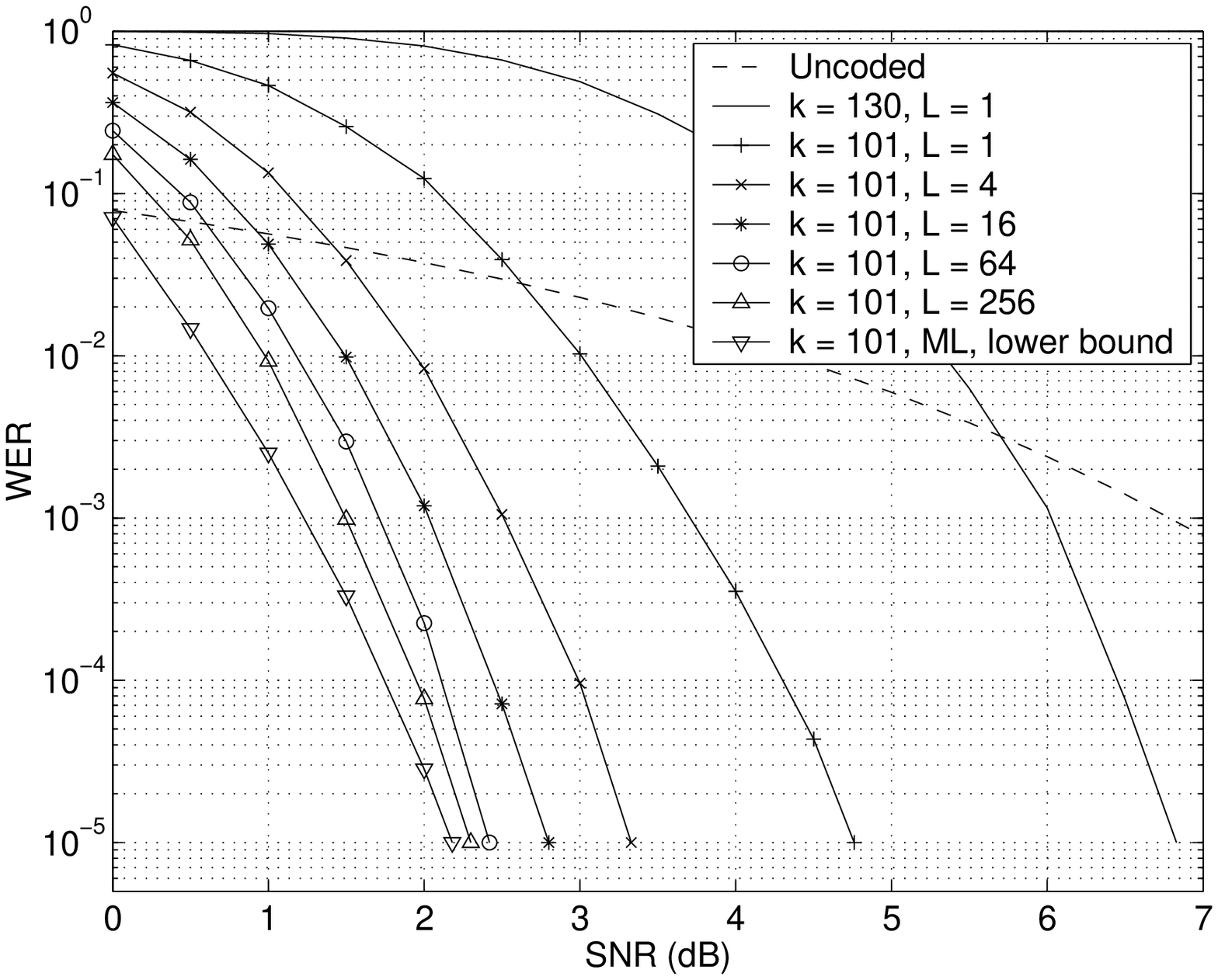}
\end{center}
\caption{($512$, $101$)-subcode of the ($512$, $130$) RM code
$\genfrac{\{}{\}}{0pt}{}{9}{3}.$ WER for the algorithm $\Psi_{r}^{m}(L)$ with
lists of size $L.$}%
\label{fig:srm0309}%
\end{figure}

In Fig. \ref{fig:srm0309}, we present similar results for the $(512,101)$%
-subcode of the $(512,130)$-code $\left\{
%TCIMACRO{\QATOP{9}{3}}%
%BeginExpansion
\genfrac{}{}{0pt}{}{9}{3}%
%EndExpansion
\right\}  .$ Here in Table 2, we also give a few list sizes $L,$ the
corresponding \ SNRs needed to reach the output WER $P=10^{-4},$ and the
complexity estimates $\left|  \Psi_{r}^{m}(L)\right|  $ counted by the number
of floating point operations. \ Similar results were also obtained for the
subcodes of other RM codes of length 512.\medskip\ \vspace{0.1in}

\begin{center}%
\begin{tabular}
[c]{|c|c|c|c|c|}\hline
$\overset{}{\underset{}{\text{List size }L}}$ & 1 & 4 & 16 & 64\\\hline
$\overset{}{\underset{}{\text{Complexity }}}$ & 7649 & 25059 & 92555 &
378022\\\hline
$\overset{}{\underset{}{\text{SNR at }10^{-4}}}$ & 4.31 & 3 & 2.5 &
2.1\\\hline
\end{tabular}
\medskip
\end{center}

Table 2. ($512$, $101$)-subcode of the ($512$, $130$) RM code $%
%TCIMACRO{\QATOPD{\{}{\}}{9}{3}}%
%BeginExpansion
\genfrac{\{}{\}}{0pt}{}{9}{3}%
%EndExpansion
.$ List sizes $L,$ the corresponding SNRs$,$ and complexity estimates $\left|
\Psi_{r}^{m}(L)\right|  $ needed at WER $10^{-4}.$\medskip

These simulation results show that combining both techniques - eliminating the
least protected bits and using small lists of codewords - gives a gain of 3 to
4 dB on the lengths $n\leq512$ over the original non-list decoding algorithm
$\Psi_{r}^{m}.$ For subcodes, we also approach ML decoding with the lists
reduced 4 to 8 times relative to the original RM codes.

\subsection{New permutation techniques}

The second improvement to the algorithm $\Psi_{r}^{m}(L)$ utilizes the rich
\textit{symmetry group} $GA(m)$ of RM codes \cite{MS} that includes
$2^{O(m^{2})}$ permutations of $n$ positions $i=(i_{1},\ldots,i_{m}).$ To
employ fewer permutations, we first\textit{ }permute the $m$ indices
$(1,2,...,m)$ of all $n$ positions $i=(i_{1},\ldots,i_{m}).$ Thus, we first
take a permutation%
\[
(1,2,...,m)\overset{\pi}{\mapsto}\left(  \pi(1),\ldots,\pi(m)\right)
\]
of $m$ \textit{indices} and consider the corresponding $m!$ permutations
$\pi(i)$ of positions $i:$
\begin{equation}
\pi(i):(i_{1},\ldots,i_{m})\rightarrow(i_{\pi(1)},\ldots,i_{\pi(m)}).
\label{perm}%
\end{equation}

\textit{Remark.} Note that the $m$ indices represent the different
\textit{axes} in $E_{2}^{m}.$ Thus, any permutation of indices is the
permutation of axes of $E_{2}^{m}.$ For example, the permutation
$(2,1,3,4,...,m)$ of $m$ indices leaves unchanged the first and the fourth
quarters of all positions $1,...,n,$ but changes the order of the second and
the third quarters.

Given a permutation $\pi,$ consider the subset of $r$ original indices\ (axes)
$\pi^{-1}\left\{  1,...,r\right\}  $ that were transformed into the first $r$
axes $1,...,r$ by the permutation $\pi$. We say that two permutations $\pi$
and $\eta$ are equivalent if these images form the identical (unordered)
subsets
\[
\pi^{-1}\left\{  1,...,r\right\}  =\eta^{-1}\{1,...,r\}.
\]
Now consider any subset $T$ of permutations (\ref{perm}) that includes exactly
one permutation from each equivalent class. Thus, $T$ includes $\left(
_{\,r}^{m}\right)  $ permutations, each of which specifies a subset of the
first \ $r$ indices. Recall that these $r$ indices correspond to the axes that
are processed first on the subpath $0^{r}$ (for example, we can start with the
axis $i_{2}$ instead of $i_{1},$ in which case we first fold the adjacent
quarters instead of the halves of the original block). Thus, this subset $T$
specifies all possible ways of choosing $r$ unordered axes that will be
processed first by the algorithm $\Psi_{r}^{m}.$

Given some positive integer $l$ (which is smaller than the former parameter
$L),$ we then incorporate these permutations $\pi(i)$ into the list decoding
$\Psi_{r}^{m}(l).$ Namely, \ we form all permutations $\mathbf{y}_{\pi(i)}$ of
the received vector $\mathbf{y}$ and apply algorithm $\Psi_{r}^{m}(l)$ to each
vector $\mathbf{y}_{\pi(i)}$. However, at each step of the algorithm, we also
\textit{combine different} lists and leave only $l$ best candidates in the
combined list, each counted once.

Note that this technique makes only marginal changes to our conventional list
decoding $\Psi_{r}^{m}(l).$ Indeed, the single vector $\mathbf{y}$ \ in our
original setting (\ref{ini1}) is replaced by $\left(  _{\,r}^{m}\right)
$\ permutations $\mathbf{y}_{\pi(i)}.$ Thus, we use parameter $\left(
_{\,r}^{m}\right)  $ in our initial setting but keep parameter $l$ for all
decoding steps. If $l<\left(  _{\,r}^{m}\right)  ,$ then the number of records
drops to $l$ almost immediately, after the first decoding is performed on the
path $0^{r}.$

Also, information bits are now decoded in different orders depending on a
specific permutation $\pi(i).$ Note that we may (and often do) get the same
entries repeated many times. Therefore, in steps 2 and 3 we must eliminate
identical entries. This is done in all steps by applying inverse permutations
and comparing the corresponding blocks $\mathbf{a}$. This permutation-based
algorithm is called $\Upsilon_{r}^{m}(l)$ below and has complexity similar to
$\left|  \Psi_{r}^{m}(l)\right|  $ for all the codes tested.

The motivation for this algorithm is as follows. The specific order of our
axes also defines the order \ in which the decoding algorithm folds\ the
original block into the subblocks of lengths $n/2,$ then $n/4,$ and so on. Now
note that this folding procedure will likely accumulate the errors whenever
erroneous positions substantially disagree on the two halves (correspondingly,
quarters, and so on). This can also happen if the errors are unevenly spread
over the two halves of the original block. By using many permutations, we make
it more likely that the error positions are spread more evenly even if they
get accumulated in the original setting $\pi(i)=i$ or any other specific
setting. In this way, \ permutation techniques serve the same functions as
interleaving does on the bursty channels.

\begin{figure}[tbh]
\begin{center}
\includegraphics[width=3.5in]{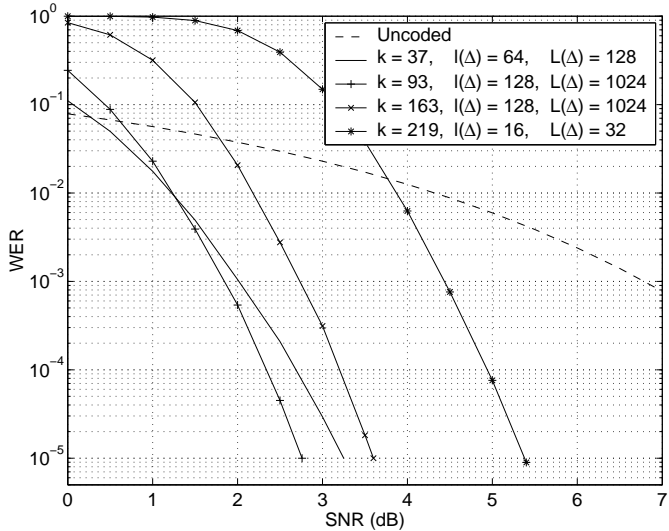}
\end{center}
\caption{Tight lower bounds on WER of ML decoding for four RM codes of length
256. The legend gives the list sizes $l(\Delta)$ and $L(\Delta)$ for which the
algorithms $\Upsilon_{r}^{m}(l)$ and $\Psi_{r}^{m}(L)$ perform within
$\Delta=0.25$ dB from these bounds.}%
\label{fig:rm256}%
\end{figure}

Simulation results for the moderate lengths 256 and 512 show that the
algorithm $\Upsilon_{r}^{m}(l)$ approaches the optimal ML performance even
when the combined list of $l$ most probable candidates \textit{is reduced two
to eight times} relative to the original algorithm $\Psi_{r}^{m}(L)$. For RM
codes of length 256, we summarize these results in Fig.~\ref{fig:rm256}. For
each RM code, we first present the lower bounds for the ML decoding error
probability. Similarly to Fig.~\ref{fig:rm128}, we then find the minimum size
$l(\Delta)$ that makes the algorithm $\Upsilon_{r}^{m}(l)$ perform only within
$\Delta=0.25$ dB away from ML decoding$.$ These sizes and complexity estimates
$\left|  \Upsilon_{r}^{m}(l)\right|  $ are also given in Table 3$.$ Note that
both algorithms give smaller lists once this performance loss $\Delta$\ is
slightly increased. In particular, the results in Table 4 show that the lists
are reduced two times for $\Delta=0.5$ dB. \

In summary, the permutation algorithm\textit{ }$\Upsilon_{r}^{m}(l)$ \textit{
performs within 0.5 dB from ML decoding on the length 256,} by processing
\ $l\leq64$ vectors for all RM codes.\textit{ }To date,\textit{ }both
techniques - permutation decoding $\Upsilon_{r}^{m}(l)$\ of complete RM codes
and list decoding $\Psi_{r}^{m}(L)$ of their subcodes - yield the best
trade-offs between near-ML performance and its complexity known on the lengths
$n\leq256.$\medskip

\begin{tabular}
[c]{|c|c|c|c|c|}\hline
RM Code & $%
%TCIMACRO{\QATOPD{\{}{\}}{8}{2}}%
%BeginExpansion
\genfrac{\{}{\}}{0pt}{}{8}{2}%
%EndExpansion
$ & $%
%TCIMACRO{\QATOPD{\{}{\}}{8}{3}}%
%BeginExpansion
\genfrac{\{}{\}}{0pt}{}{8}{3}%
%EndExpansion
$ & $\underset{}{\overset{}{%
%TCIMACRO{\QATOPD{\{}{\}}{8}{4}}%
%BeginExpansion
\genfrac{\{}{\}}{0pt}{}{8}{4}%
%EndExpansion
}}$ & $%
%TCIMACRO{\QATOPD{\{}{\}}{8}{5}}%
%BeginExpansion
\genfrac{\{}{\}}{0pt}{}{8}{5}%
%EndExpansion
$\\\hline
$\overset{}{\underset{}{\text{List size }l(\Delta)}}$ & 64 & 128 & 128 &
16\\\hline
$\underset{}{%
\begin{array}
[c]{c}%
\text{Complexity }\\
\left|  \Upsilon_{r}^{m}(l)\right|
\end{array}
}$ & 216752 & 655805 & 777909 & 94322\\\hline
$\overset{}{\underset{}{\text{SNR at }10^{-4}}}$ & 2.91 & 2.65 & 3.38 &
5.2\\\hline
\end{tabular}
\medskip

Table 3. RM codes of length 256: \ the list sizes, complexities, and the
corresponding SNRs, at which the permutation algorithm $\Upsilon_{r}^{m}(l)$
performs within $\Delta=0.25$ dB from ML decoding at WER $10^{-4}.$\medskip%

\begin{tabular}
[c]{|c|c|c|c|c|}\hline
RM Code & $%
%TCIMACRO{\QATOPD{\{}{\}}{8}{2}}%
%BeginExpansion
\genfrac{\{}{\}}{0pt}{}{8}{2}%
%EndExpansion
$ & $%
%TCIMACRO{\QATOPD{\{}{\}}{8}{3}}%
%BeginExpansion
\genfrac{\{}{\}}{0pt}{}{8}{3}%
%EndExpansion
$ & $\underset{}{\overset{}{%
%TCIMACRO{\QATOPD{\{}{\}}{8}{4}}%
%BeginExpansion
\genfrac{\{}{\}}{0pt}{}{8}{4}%
%EndExpansion
}}$ & $%
%TCIMACRO{\QATOPD{\{}{\}}{8}{5}}%
%BeginExpansion
\genfrac{\{}{\}}{0pt}{}{8}{5}%
%EndExpansion
$\\\hline
$\overset{}{\underset{}{\text{List size }l(\Delta)}}$ &
\multicolumn{1}{|c|}{32} & 64 & 64 & 8\\\hline
$\underset{}{%
\begin{array}
[c]{c}%
\text{Complexity }\\
\left|  \Upsilon_{r}^{m}(l)\right|
\end{array}
}$ & 116471 & 333506 & 389368 & 37756\\\hline
$\overset{}{\underset{}{\text{SNR at }10^{-4}}}$ & 3.12 & 2.82 & 3.55 &
5.4\\\hline
\end{tabular}
\medskip

Table 4. RM codes of length 256: the\ list sizes, complexities, and the
corresponding SNRs, at which the permutation algorithm $\Upsilon_{r}^{m}(l)$
performs within\ $\Delta=0.5$ dB from ML decoding at WER $10^{-4}.$
%\end{onecolumn}
%\bigskip\newpage
\medskip

Note, however, that the algorithm $\Upsilon_{r}^{m}(l)$ gives almost no
advantage for the subcodes considered in the previous subsection. Indeed,
these subcodes are obtained by eliminating the leftmost (least protected)
information bits. However, any new permutation $\pi(i)$ assigns the new
information bits to these leftmost nodes. Thus, the new bits also become the
least protected. Another unsatisfactory observation is that increasing the
size of the permutation set $T$ - say, to include all $m!$ permutations of all
$m$ indices - helps little in improving decoding performance. More generally,
there are a number of important open problems related to these permutation
techniques. We name a few:

$-$ find the best permutation set $T$ for the algorithm $\Upsilon_{r}^{m}(l)$;

$-$ analyze the algorithm $\Upsilon_{r}^{m}(l)$ analytically;

$-$ modify the algorithm $\Upsilon_{r}^{m}(l)$ for subcodes.

\section{Concluding remarks}

In this paper, we considered recursive decoding algorithms for RM codes that
can provide near-maximum likelihood decoding with feasible complexity for RM
codes or their subcodes on the moderate lengths $n\leq512$.

Our study yet leaves many open problems. Firstly, we need to tightly estimate
the error probabilities $p(\xi)$ on the different paths $\xi.$ To optimize our
pruning procedures for specific subcodes, it is important to find the order in
which information bits should be removed from the original RM code. Finally,
it is yet an open problem to analytically estimate the performance of the
algorithms $\Psi_{\,r}^{m}(L)$ and $\Upsilon_{r}^{m}(l).$

\end{document}